\documentclass[copyright,creativecommons]{eptcs}
\providecommand{\event}{ThEdu 2022}

\usepackage{underscore}
\usepackage[T1]{fontenc}
\usepackage{subcaption}
\usepackage{graphicx}
\usepackage{amsmath}
\usepackage{amssymb}
\usepackage{cite}
\usepackage{cleveref}

\usepackage{isabelle, isabellesym}
\isabellestyle{it} 
\newcommand{\SNIP}[2]{\expandafter\newcommand\csname snippet--#1\endcsname{#2}}

\newcommand{\GetSnip}[1]{%
    \ifcsname snippet--#1\endcsname%
        \csname snippet--#1\endcsname%
    \else%
        \PackageWarning{snips}{Snippet ``#1'' is undefined.}%
        \emph{Warning: Snippet ``#1'' is undefined.}%
    \fi%
}

\newcommand{\Snippet}[1]{{%
  \newcount\i
  \i=0
  \loop
    \GetSnip{#1-\the\i}
    \advance \i 1
  \ifcsname snippet--#1-\the\i\endcsname
  \repeat
}}

\newcommand{\SnippetPart}[3]{{%
  \newcount\i
  \i=#1
  \loop
    \ifnum \i=#2
      \renewcommand{\isanewline}{}%
    \fi
    \GetSnip{#3-\the\i}
    \advance \i 1
    \ifnum \i>#2 {}
    \else \repeat
}}

\RequirePackage{array}
\newenvironment{authors}[1]%
  {\begingroup
   \newcommand\estyle{}%
   \renewcommand\institute[1]%
     {\\\multicolumn{#1}{@{}c@{}}{\scriptsize\begin{tabular}[t]{@{}>{\footnotesize}c@{}}##1\end{tabular}}}%
   \renewcommand\email[1]%
     {\gdef\estyle{\footnotesize\ttfamily}\\##1\gdef\estyle{}}
   \begin{tabular}[t]{@{}*{#1}{>{\estyle}c}@{}}
  }%
  {\end{tabular}%
   \endgroup
  }

\title{On Exams with the Isabelle Proof Assistant}
\author{
  \begin{authors}{2}
    Frederik Krogsdal Jacobsen & Jørgen Villadsen
      \institute{Technical University of Denmark, Kongens Lyngby, Denmark}
      \email{fkjac@dtu.dk & jovi@dtu.dk}
  \end{authors}
}

\def\authorrunning{F. K. Jacobsen \& J. Villadsen}
\def\titlerunning{On Exams with the Isabelle Proof Assistant}

\begin{document}
\maketitle

\begin{abstract}
We present an approach for testing student learning outcomes in a course on automated reasoning using the Isabelle proof assistant.
The approach allows us to test both general understanding of formal proofs in various logical proof systems and understanding of proofs in the higher-order logic of Isabelle/HOL in particular.
The use of Isabelle enables almost automatic grading of large parts of the exam.
We explain our approach through a number of example problems, and explain why we believe that each of the kinds of problems we have selected are adequate measures of our intended learning outcomes.
Finally, we discuss our experiences using the approach for the exam of a course on automated reasoning and suggest potential future work.
\end{abstract}

\section{Introduction}
At the Technical University of Denmark, we currently teach a MSc level course on automated reasoning using the Isabelle proof assistant \cite{Isabelle02} as our main tool.
The course is a 5 ECTS optional course and the homepage is here:

\begin{center}
\url{https://courses.compute.dtu.dk/02256/}
\end{center}

The course is an introduction to automatic and interactive theorem proving, and Isabelle is used to formalize almost all of the concepts we introduce during the course.
We have developed a number of external tools to allow us to teach basic proofs in  natural deduction and sequent calculus while slowly progressing towards showing students the full power of Isabelle.
The learning objectives of the course are as follows:
\begin{enumerate}
    \item explain the basic concepts introduced in the course
    \item express mathematical theorems and properties of IT systems formally
    \item master the natural deduction proof system
    \item relate first-order logic, higher-order logic and type theory
    \item construct formal proofs in the procedural style and in the declarative style
    \item use automatic and interactive computer systems for automated reasoning
    \item evaluate the trustworthiness of proof assistants and related tools
    \item communicate solutions to problems in a clear and precise manner
\end{enumerate}

We expect students to already know some logic and to be relatively proficient in functional programming before starting the course.
Additionally, we expect students to have some basic knowledge of artificial intelligence algorithms for deduction.
Our undergraduate program in computer science and engineering, which many of our students have completed, contains several courses that introduce students to these topics.
Many of our MSc students are however from other universities or from other undergraduate programs at our university, and may thus only be familiar with some of the topics we consider prerequisites.
At our institution, prerequisites are not mandatory, but only recommended (partly to avoid a large administrative burden of attempting to determine equivalences between our undergraduate courses and courses at other universities).
For this reason, we cannot expect students to be familiar with all prerequisites, and we thus endeavour to explain the prerequisites throughout the course.

During the course we need to test student learning outcomes.
We ask students to hand in six assignments during the course, but we also have a two hour written exam at the end of the course.
In the present paper, we will explain how we have designed this exam to make use of the automation afforded by Isabelle while testing understanding of both logic in the general sense and the specifics of proving theorems in Isabelle.
We find that this setup works well for our course, which had 41 registered students at the time of the exam.

The exam is a two hour written exam with all aids allowed (including internet access, though students are of course not allowed to ask others for help during the exam).
The exam sheet consists of an Isabelle file to be filled in as well as a ``library'' file containing the definitions needed for the proof systems used in the exam.
The exam consists of five problems (weighted equally), each of which contains two questions.
The problems in the exam can be solved in any order.
The five problems concern the following topics:
\begin{enumerate}
    \item Isabelle proofs without automation
    \item Verification of functional programs in Isabelle/HOL
    \item Natural deduction proofs
    \item Sequent calculus proofs
    \item General proofs in Isabelle/HOL with Isar
\end{enumerate}

In the present paper, we will give examples of the kinds of questions we have designed for each topic, and explain why we believe these questions are adequate tests of learning outcomes for each topic.
Each problem is weighted equally, but the two questions within each problem may have different weights.
This allows us to create one easier and one harder question in some of the problems, which we find helps students not to ``freeze'' during the exam by giving them some early successes.

The examples and solutions given in this paper are drawn from a test exam which is also handed out to students so they can practice the format (with no impact on their grade) before the actual exam.
At our institution, students are usually given full access to exam questions from previous years, often including solutions.
Since we changed the format of the exam quite significantly this year, we introduced the test exam to give students a chance to ask questions about how to fill in the Isabelle file they were given, and to let students see the types of questions they would be asked in the actual exam.
It should be noted that the exam problems mentioned in the present paper do not test all of the learning objectives mentioned above.
The assignments handed in during the course test the remaining learning objectives and are also considered when giving the final grade.

One of the main benefits of using Isabelle for our exam is of course that much of the grading is almost automatic.
Since Isabelle can check whether the proofs handed in by the students are correct, all that is needed to grade the exam are a few manual checks for style and to make sure that the students have not changed any definitions such that they are proving something different than what they were asked to.
We have not experienced students maliciously changing definitions to make proofs easier, but some students misunderstand the assignment and write faulty proofs, then change definitions (in an obvious, non-obfuscated way) to make the proofs go through.
For larger courses it may be desirable or necessary to completely automate the grading, but doing so puts much stricter constraints on the kinds of questions we can ask, and we have thus chosen not to do this.
For instance, it becomes impossible to give partial points to students who have the right proof idea, but hand in files with minor syntax errors (e.g.\ forgetting to end their otherwise correct proofs with a \isa{qed}), which happens quite often.
Some of the challenges in using the fully automatic approach have been described by Pierce~\cite{Pierce09}.
For smaller courses, we believe that an oral exam or an exam based on a project may be both easier to set up and more beneficial, but at the scale of our course we simply do not have the time for these.

We also teach a BSc level course on logical systems and logic programming, which provides the prerequisite logic for many of our students.
In that course the students are briefly introduced to Isabelle, but only in a much more limited setting, and they do not use Isabelle during the final exam \cite{FMTea}.

In the next section, we will compare our approach to a selection of related work.
In the sections after that, we will discuss and showcase some of the questions and solutions and explain which learning outcomes they test.
We will then describe our experiences using the exam questions in practice and discuss possible future work before concluding.

\section{Related work}
The proof assistant Isabelle/HOL has previously been used to teach programming language semantics and their applications at MSc level \cite{Nipkow-LSD}.
This course specifically avoids focusing on logic itself, and instead uses both logic and Isabelle as tools to understand semantics.
In contrast, the aim of our course is exactly to teach logic, and thus also parts of the inner workings of Isabelle.
Additionally, our course does not focus very much on applications.
On the other hand, the courses are similar in that we also do not focus on teaching the proof assistant for its own sake, but rather as a tool to understand and use logic.
Importantly, the semantics course does not use Isabelle at all for the exam.

More recently, Isabelle/HOL has also been used to teach algorithms and data structures \cite{Nipkow-CPP}.
This course also does not focus on logic itself, but uses logic and Isabelle as tools to understand algorithms and data structures.
Some introduction to Isabelle is however necessary, and this takes up the first third of the course.
This course also uses Isabelle for the exam, but does not focus on logic.

The proof assistant Coq has been used for a broad series of textbooks on basic logic, programming languages, functional algorithms and separation logic \cite{Pierce:SF1}.
These books also focus quite a bit on learning the Coq proof assistant, the user interface of which is further removed from pen-and-paper proofs than the interface of Isabelle.
The books feature many exercises which are all to be completed in Coq, as well as scripts that can automatically grade most of the exercises.
The exercises have been specifically designed such that they can be graded automatically.
We believe that this is more easily done for proofs about the topics of programming languages and algorithms than for proofs about logic, since proofs about logic are, in our experience, more difficult to split into independent lemmas and require more auxiliary concepts.
The books only cover basic logic, and mostly to demonstrate features of Coq.
The exams in the courses they have been used in do not, as far as we know, use Coq, though homework exercises in Coq have been used \cite{Pierce09}.

The proof assistant Lean has been used to teach programming language semantics and the basic foundations of mathematics \cite{LeanCourse2022}.
The course also focuses quite a bit on learning the Lean proof assistant itself.
The exam of this course is on paper, but students are given problems using the Lean syntax and are also expected to write several of their answers using Lean syntax.
The exam includes questions about the types of certain Lean terms, questions asking the students to write definitions in Lean, and questions asking students to prove certain properties of definitions in Lean.
The proofs do not have to be actual Lean proofs, but do need to be very formal.

Lean has also been used to teach mathematics in general \cite{NNG2021, LeanCourse2015}.
A longer review of the use of proof assistants for general mathematics is available in \cite{Avigad2014}.

The proof assistant Agda has been used to teach logic and functional programming in a course at our institution, but only for a single example, and not for the exam \cite{TFPIE}.

\section{Isabelle proofs without automation}
The first problem of the exam is designed to test whether the student understands the proof system of higher-order logic, which is the underpinning of Isabelle/HOL.
We do this by asking students to prove the validity of simple logical formulas such as $p \leftrightarrow \neg \neg p$ and
\begin{equation*}
    (\exists x. \forall y. r(x, y)) \rightarrow (\forall y. \exists x. r(x, y))
\end{equation*}
The students must formally prove the validity of the formulas within Isabelle, and this problem thus also tests whether students master the natural deduction proof system used for Isabelle proofs and whether students actually know how to use Isabelle.

To avoid students simply using Isabelle's automated tools such as Sledgehammer to prove the formulas, we provide a re-implementation of higher-order logic without any automation within Isabelle/Pure, and require students to use the proof rules from this system directly.
We have recently described this re-implementation for intuitionistic and classical propositional logic \cite{ThEdu21}.
For the course we have added the usual quantifiers to obtain first-order logic and finally we re-implement higher-order logic \cite{SLAI-HOL}.
With this system, Isabelle can still often suggest an appropriate proof rule to use if the student first writes their intended subgoal, but only for single steps at a time, and not in all cases.

\begin{figure}
\includegraphics[width=\textwidth]{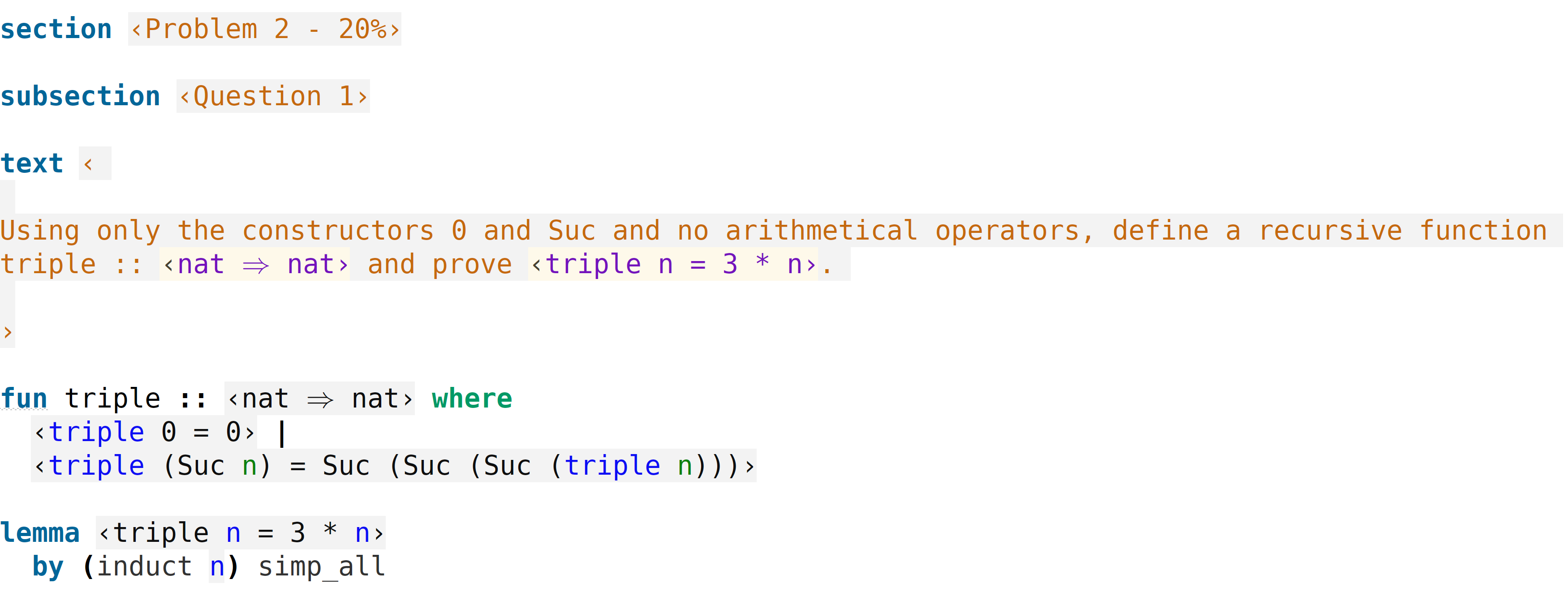}
\caption{Problem 2 --- The given text and the solution below.}%
\label{fig:P2}
\end{figure}

\section{Verification of functional programs in Isabelle/HOL}\label{sec:verification}
The second problem of the exam is mainly designed to test whether students can program and prove in Isabelle/HOL, as well as whether the students can express properties of IT systems formally.
We do this by asking students to implement very simple programs and prove simple properties of them.
These problems are similar to some of the simpler exercises in the ``Programming and Proving in Isabelle/HOL'' tutorial \cite{prog-prove}.
An example question is given in \cref{fig:P2}, which also contains a solution.

In designing questions such as these, it is quite difficult to obtain a reasonable amount of complexity in the programs and properties we ask students to write and prove.
Even very simple properties can be very difficult to prove formally, and students have only a few minutes to solve the problem under the high pressure of the exam situation.
We thus keep the programs and proofs very simple, aiming to design questions that are trivial for experienced Isabelle users, but which will be difficult to solve within a few minutes for someone who has never used Isabelle before.
Note that the solution requires the use of induction, which means that the Sledgehammer tool can not be used to prove the property directly.
The intention is that students who have actually followed the course and know how to use Isabelle should find the questions relatively easy, while any students who have not paid attention during the course should find the questions very difficult.

While the questions in the exam itself are quite easy, students are given a number of assignments with more complex questions during the course.
Topics include programming with relations and sets, defining inductive types and recursive functions over them, and proving properties by structural induction, including rule induction over inductively defined relations.

\section{Natural deduction and sequent calculus proofs}
The next two problems of the exam involve proofs in a natural deduction system and a sequent calculus system, respectively.
Both of these systems concern classical first-order logic with functions, but are simpler to use than Isabelle, which can be quite complicated.
We have designed two external tools which allow students to write formal proofs in these proof systems, which are then automatically translated to Isabelle proofs.
The natural deduction system, NaDeA, has a graphical front-end, while the sequent calculus system, SeCaV, has a textual front-end  \cite{nadea,LSFA}.
Both of these front-ends are implemented as web applications, which students can access during the exam since they have access to the internet:

\begin{center}
\url{https://nadea.compute.dtu.dk/}
\\[2ex]
\url{https://secav.compute.dtu.dk/}
\end{center}

The main idea behind these systems is to make it easier for students to understand which options they have when attempting to prove a formula valid.
In NaDeA, the graphical interface means that students cannot input formulas with syntax mistakes.
Additionally, NaDeA automatically filters the proof rules in the system such that only rules that can actually be applied in a given situation are shown in the graphical interface.
Finally, the system handles the actual application of the proof rules automatically, including keeping track of branches in the proof tree.

The SeCaV tool is closer to apply-style proofs in Isabelle, except that the proof system is sequent calculus instead of natural deduction.
The tool takes input as text, which means that students can make syntactical mistakes when inputting formulas and proof rules just as in Isabelle.
The other main difference to NaDeA is that students must manually determine what applying a proof rule actually does and write out the result of applying the rule themselves.
Since the proof system is much simpler than the proof system of Isabelle/HOL, however, the SeCaV system is able to give quite detailed warnings and error messages when students make mistakes.

We have described the pedagogical and logical design considerations of these systems in more detail elsewhere \cite{ThEdu20}.
A key point is that we have a formalization of the syntax, semantics and proof systems for both NaDeA and SeCaV.
We also have formal soundness and completeness theorems for the proof systems.
The formalizations are part of the teaching materials during the course.
In the exam, we ask students to prove both propositional and first-order formulas, and involve classical formulas such that students will need to understand how to reason in classical logic.
These problems thus test whether the student masters natural deduction, whether the student understands when classical proof rules are needed and how to use them, and whether the student can construct formal proofs using interactive computer systems.
Both NaDeA and SeCaV run in the browser of each student independently, and any number of students can thus use the tools at the same time without performance issues.

\section{General proofs in Isabelle/HOL with Isar}
\begin{figure}
\begin{subfigure}[t]{0.50\textwidth}
\includegraphics[width=\textwidth]{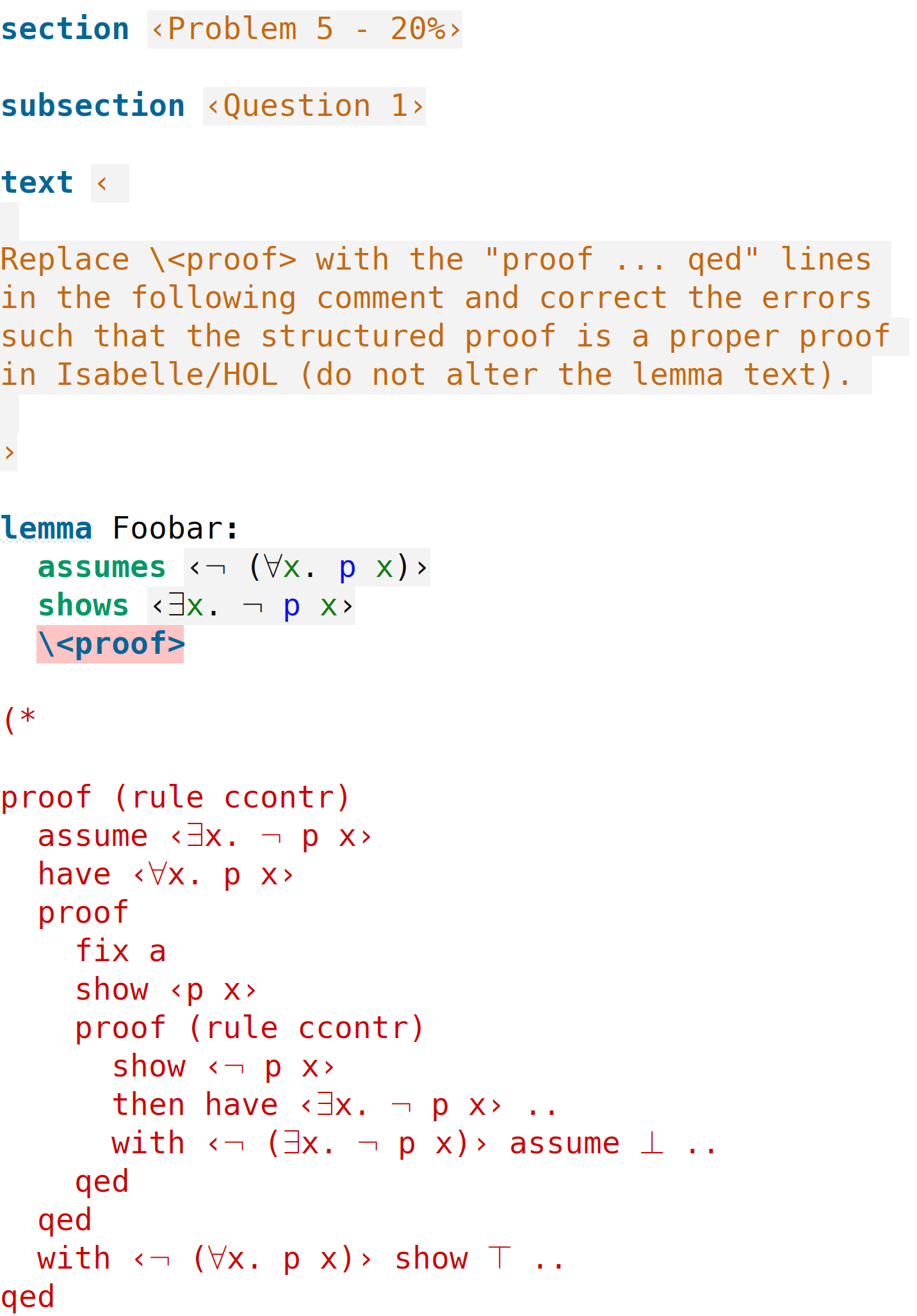}
\caption{Problem 5}%
\label{fig:P50}
\end{subfigure}%
\begin{subfigure}[t]{0.50\textwidth}
\includegraphics[width=\textwidth]{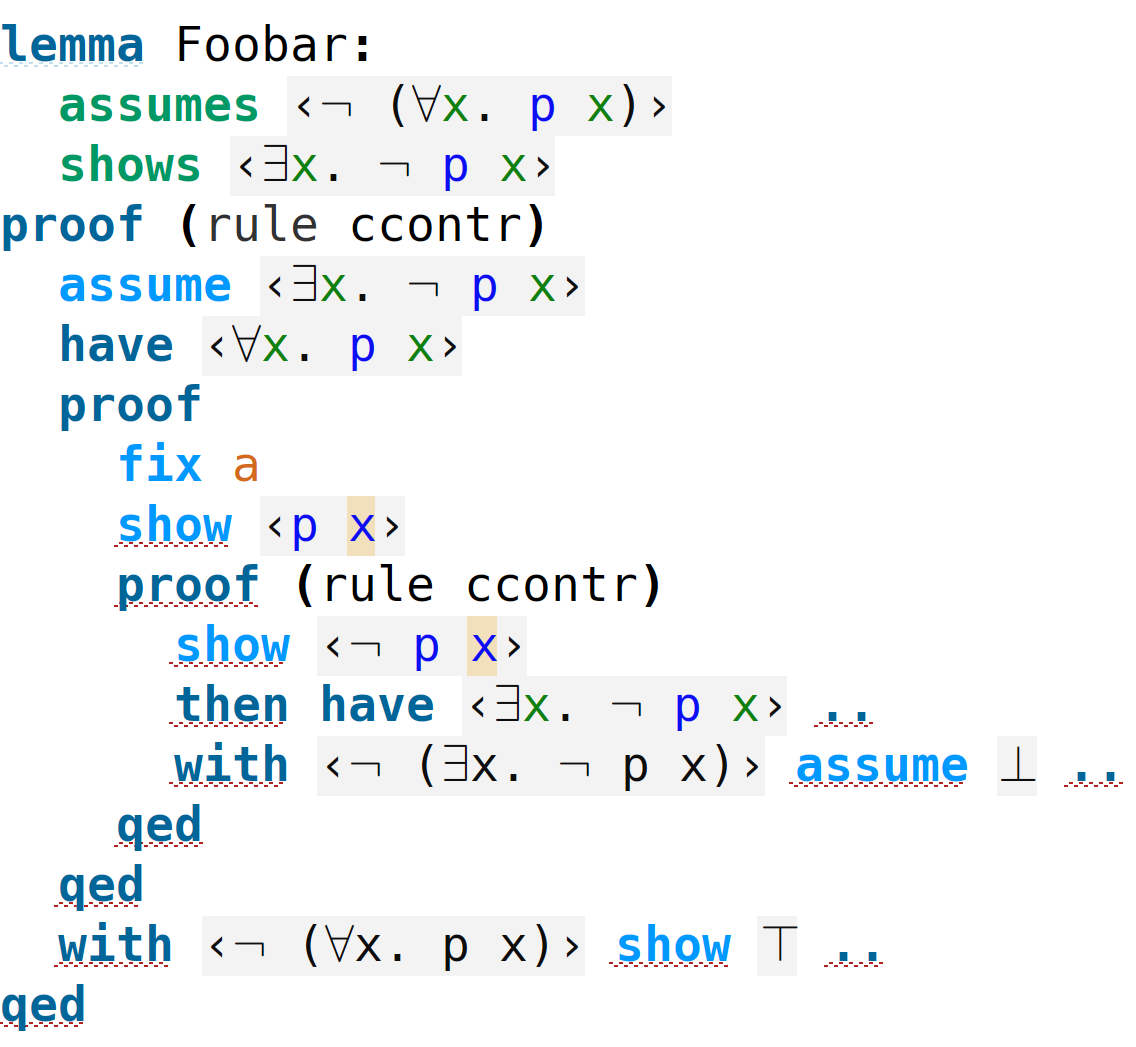}
\caption{Problem 5 --- Step 1}%
\label{fig:P51}
\end{subfigure}\\[3ex]
\begin{subfigure}[t]{0.50\textwidth}
\includegraphics[width=\textwidth]{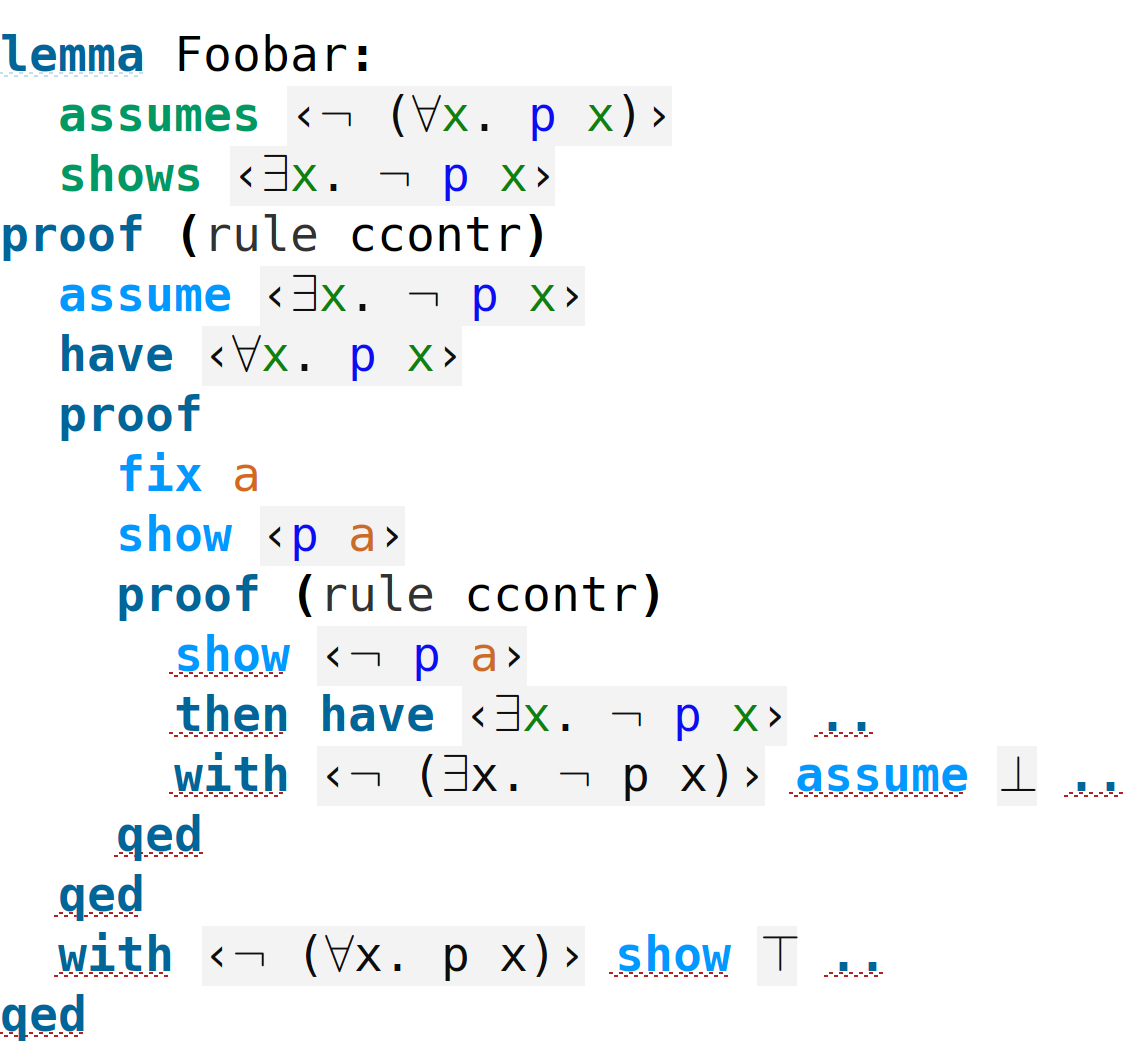}
\caption{Problem 5 --- Step 2}%
\label{fig:P52}
\end{subfigure}%
\begin{subfigure}[t]{0.50\textwidth}
\includegraphics[width=\textwidth]{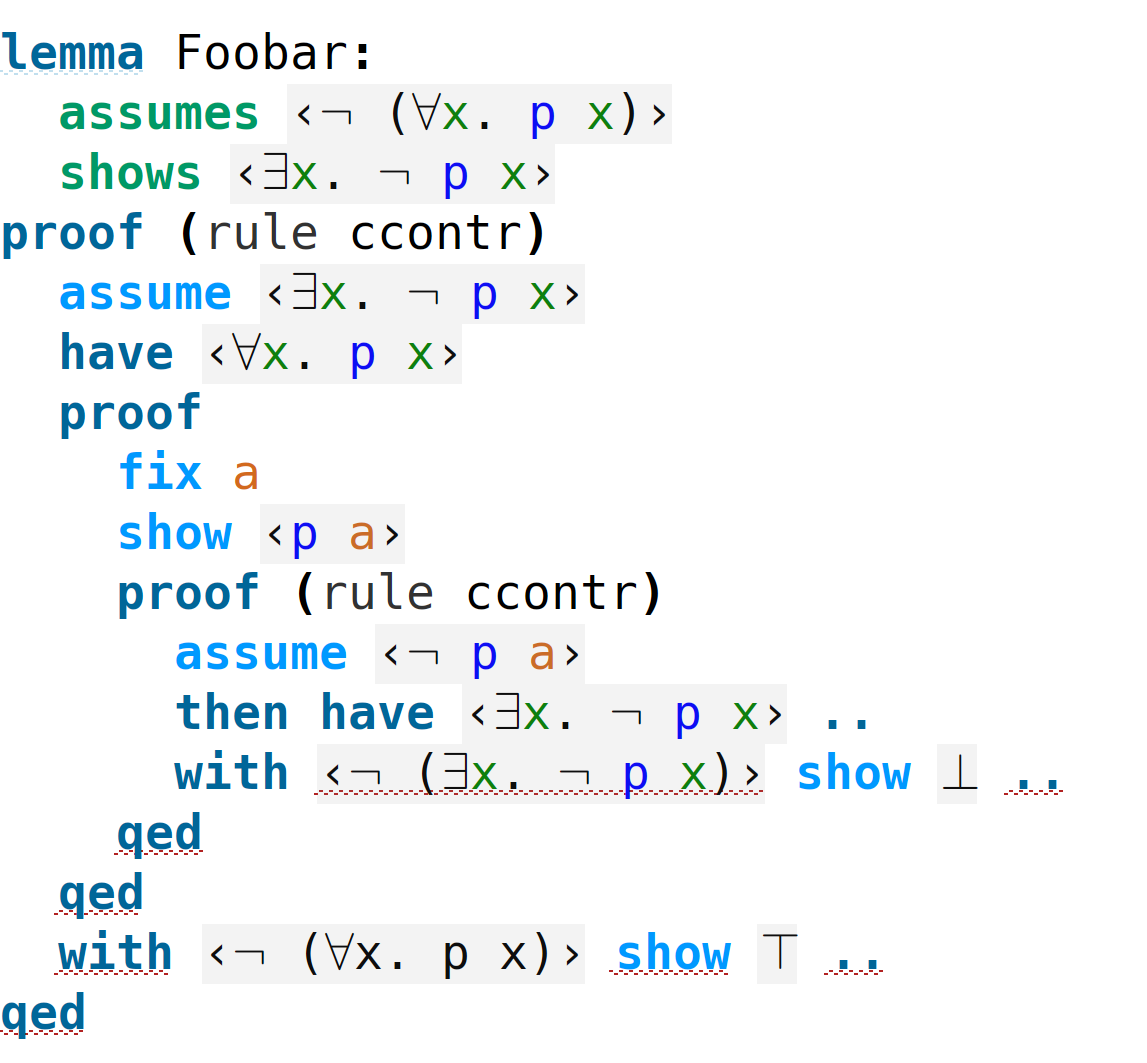}
\caption{Problem 5 --- Step 3}%
\label{fig:P53}
\end{subfigure}%
\medskip
\caption{Problem 5 --- The given text and steps 1--3.}%
\label{fig:P5-0-3}
\end{figure}

The final problem of the exam is designed to test student understanding of structured proofs in the Isar language of Isabelle/HOL \cite{isar-ref}, i.e.\ whether the student can construct formal proofs in the declarative style.
The questions in this problem ask students to finish a proof of a more complicated property, e.g.\ the one seen in \cref{fig:P50}.
The given question in \cref{fig:P50} contains a lemma and a comment with a ``proof'' of the lemma which has several mistakes.
The student must then fix these mistakes and thus finish the proof.
The idea behind this kind of question is to avoid the issue of devising properties of appropriate complexity also mentioned in \cref{sec:verification}.
By giving students a partially completed proof and asking them to fix the mistakes in it, we can ask students to prove properties that they would otherwise not be able to prove within the short time frame of the exam situation.
This allows us to test student understanding of non-trivial proofs without succumbing to merely testing that students can use the automation of Isabelle/HOL.

\begin{figure}
\begin{subfigure}[t]{0.50\textwidth}
\includegraphics[width=\textwidth]{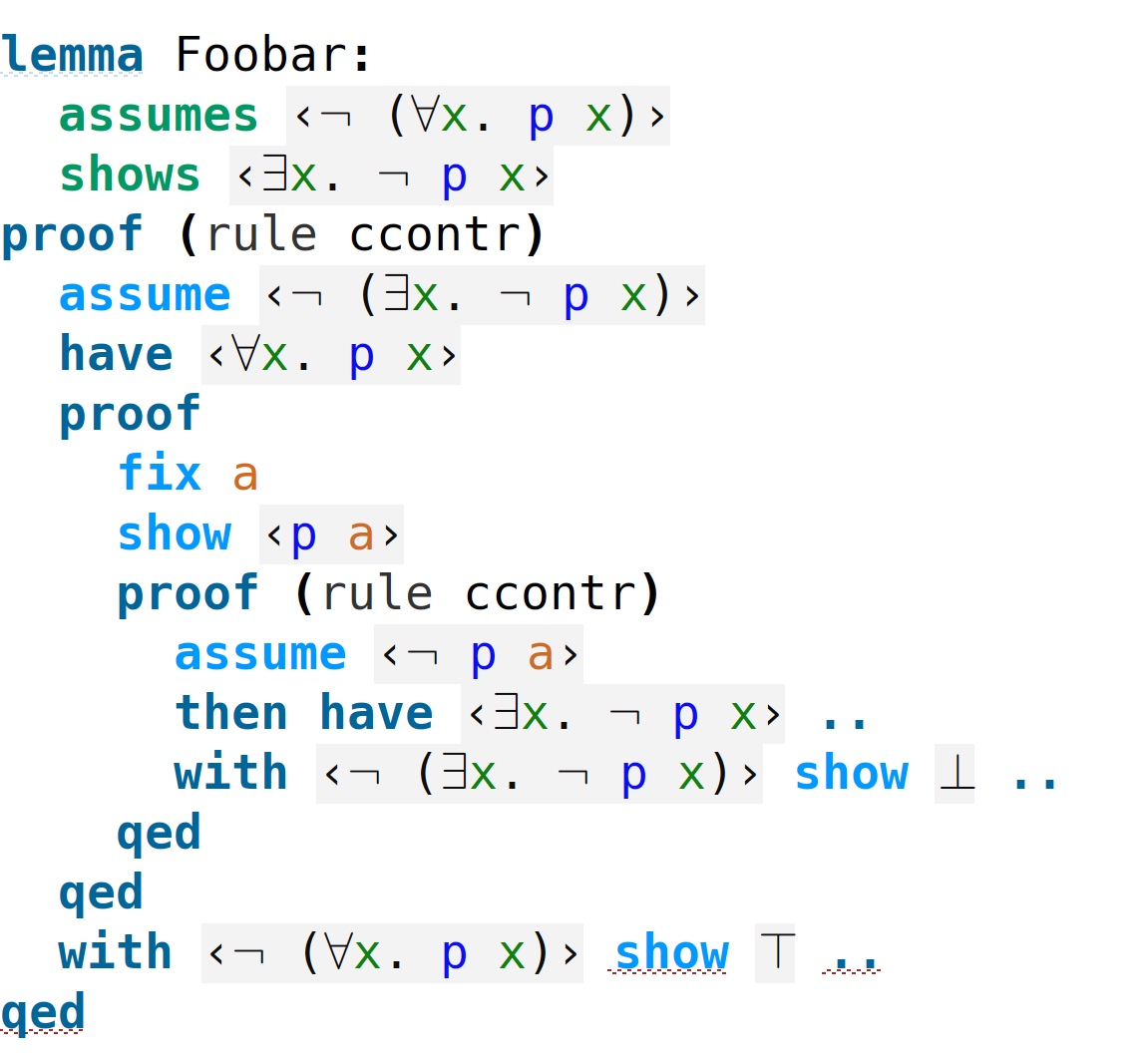}
\caption{Problem 5 --- Step 4}%
\label{fig:P54}
\end{subfigure}%
\begin{subfigure}[t]{0.50\textwidth}
\includegraphics[trim={0 -2mm 0 0},clip,width=\textwidth]{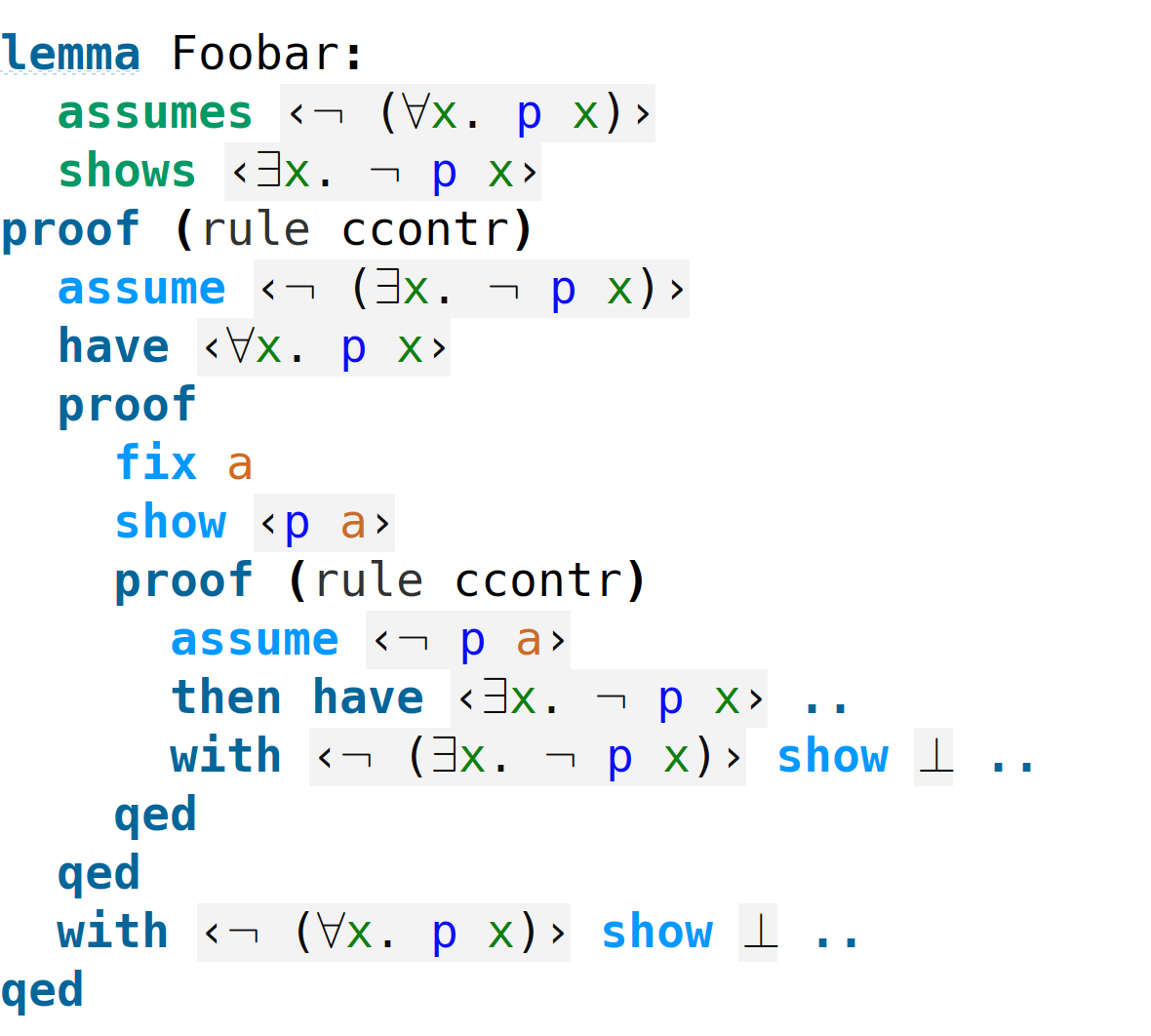}
\caption{Problem 5 --- Step 5 / Solution}%
\label{fig:P55}
\end{subfigure}%
\medskip
\caption{Problem 5 --- Steps 4--5 / Solution.}%
\label{fig:P5-4-5}
\end{figure}

We will showcase the concept by walking through a solution of a question from the test exam.
In \cref{fig:P51}, we have uncommented the partially completed proof, which induces the Isabelle/jEdit development environment to highlight the errors in the proof.
Notice the red underlining of several keywords and the brown backgrounds of some variables in the proof.
More detailed information about the errors is also available either by moving the cursor over the underlined keywords or in a separate ``Output'' pane in Isabelle/jEdit.
It should be noted that while Isabelle can highlight the errors and often provide some information about the cause, Isabelle is not able to say anything about how to fix the error.
Starting from the top, the first error we notice is that the line \isa{\isakeyword{show} \isacartoucheopen p x\isacartoucheclose} contains the variable \isa{x}, which has not been defined.
Indeed, the given proof introduces a local constant \isa{a} on the line immediately before this, so the proof can be fixed by changing every highlighted \isa{x} into an \isa{a}, as done in \cref{fig:P52}.
Next, we look at the subproof attempting to prove the goal we just fixed.
In \cref{fig:P52}, we notice that the keywords \isa{\isakeyword{show}} and \isa{\isakeyword{assume}} have been switched, so that the mistaken proof proceeds by first attempting to show the assumption, then assuming the conclusion at the end.
This is fixed by simply exchanging the keywords, which has been done in \cref{fig:P53}.
Next, the formula \isa{\isasymnot~\isacharparenleft \isasymexists x\isachardot~\isasymnot~p~x\isacharparenright}, which is used to show \isa{\isasymbottom} in the subproof, is not an assumption that is available at this point in the proof.
This error requires some more knowledge of Isabelle/HOL than the previous ones to fix, since the error is not actually on this line, but instead in the very beginning of the proof, where a wrong assumption is made.
The proof starts with the proof rule for classical contradiction, but the first assumption assumes the original formula, not its negation.
By fixing this, as is done in \cref{fig:P54}, the appropriate formula becomes available as an assumption later in the proof.
The final mistake in the proof is also related to the proof by contradiction, since the final line of the proof attempts to show \isa{\isasymtop} while the correct way to end a proof by contradiction is to show \isa{\isasymbottom}.
Fixing this we obtain \cref{fig:P55}, which Isabelle verifies as being a correct proof.

While this lemma and its proof are quite simple, the time constraints of the exam situation means that students will need to identify and fix the mistakes very quickly.
During the course, students are asked to prove more complicated lemmas from scratch during exercises and as assigned problems that are considered when giving the final grade.
For these lemmas, students have much more time, and access to teaching assistants who can help steer students onto the right track.
In the exam situation, we would like to prevent the situation where solving the problem requires ``getting the right idea'', since students will not have very much time to experiment with different approaches.
We find that providing the students with an outline in the form of a partial proof helps prevent this issue.
It would of course also be an option to give the students more time for the exam, but this would increase the workload for designing and grading the exam questions significantly, and make it harder to find space and time slots for the exam during the busy exam period.

\section{Experiences in practice}
We have used the exam problems in our course during the most recent exam in May.
Of the 41 students registered for the exam, 36 passed the course.
The remaining five students did not show up for the exam, and two additional students deregistered for the exam before the deadline.

The grades for the course were given based on both the exam and the assignments handed in during the course.
The assignments were graded throughout the course and the students were given feedback after handing in each assignment.
The final grade was given as an evaluation of the whole based on the assignments and the exam.
There was a good correlation between the results in the assignments and the results in the exam, and no student grade moved more than a single step when comparing the overall grade to the preliminary grade suggested by their performance in the assignments.

The grade average for the course was 9.9 out of 12 (in the Danish national grading system, in which 12 is equivalent to the ECTS grade A and 10 is equivalent to the ECTS grade B).
While this may seem high, it should be noted that this average only includes the students who actually showed up for the exam, and that it is common for Danish students to skip an exam if they do not feel prepared to take it, since they may then take a new exam later without ``risking'' passing with a low final grade due to lack of preparation for the exam.
Skipping an exam in this manner does not impact the grades of a student as long as the student passes the exam within 3 attempts, but students who pass the exam are not allowed to take a new exam.
If we were instead to count students who did not show up for the exam with the lowest possible grade, the average would instead be 7.8 (slightly above ECTS grade C), which is in line with the intended average of the Danish scale.
Another possible reason for the relatively high grade average is of course that our course is an advanced, non-mandatory course, which means that students self-select towards those who are motivated and interested in the topic of the course.
At our institution, it is possible for students to audit courses for the first several weeks before committing to taking the course, and indeed a bit more than one third of the 69 initial students had dropped the course by the deadline for doing so.
We presume that many of these students would have performed less well in the exam than those who stayed.

Adjusting the grades based on relative performance after seeing the exam results to obtain a specific distribution of grades (so-called ``curving'') is illegal in Denmark, and is thus not an option.
Instead, instructors are expected to adjust the difficulty of courses over several semesters to obtain the intended distribution of grades.
We will consider making the course slightly more difficult next time it runs to move towards a lower average grade.

Students did not have any issues understanding how to use Isabelle or how to hand in their solutions during the exam.
The submitted solutions also generally indicate that students understood what they were supposed to do to solve every problem, although they were of course not always able to provide a correct solution.
Many students handed in Isabelle files with syntax errors, but these were generally quite easy to fix, and consisted primarily of abandoned proof attempts not being marked as such.

We generally found the exam submissions quite easy to grade.
Correct solutions to the questions could in most cases be reviewed almost automatically by simply noticing that the student followed a reasonable approach and that Isabelle found no mistakes in their proof.
Partial solutions were harder to grade, since it is of course difficult to estimate how close the student was to actually finishing the proof.
This difficulty is no larger than it would have been for a pen and paper proof, however, and again Isabelle was of assistance by highlighting mistakes and missing parts in the proofs.

Since students had access to all aids (including the internet) during the exam, we should mention the possibility of cheating.
The exam was supervised by several proctors to detect cheating, and mobile phones were not allowed in the exam hall.
If students had cheated by communicating with each other, we would expect to see identical solutions, but we did not notice any irregularities in this regard.
It might be possible to mitigate cheating by giving students slightly different problems, but we believe it would be hard to ensure that such problems would be of the same difficulty.
We have developed versions of our auxiliary tools which can be downloaded and run locally (offline) in a browser, and in future iterations of the exams we consider not allowing students access to the internet.

In the rest of this section, we will describe the student evaluation of the exam, then discuss in more detail the performance and issues students displayed while attempting to solve each of the types of problems in our exam.

\subsection{Student evaluation}
All students at our institution are asked to anonymously evaluate exams they participate in.
The evaluation form for the exam for the course was available to 43 students (including the two students who deregistered for the exam before the deadline).
The evaluation was filled in by 17 students.
Of the students who evaluated the course, 15 had not yet received their grade at the time of evaluation, while one student had.
A single student filled in the evaluation by indicating that they had not been following the course.

The first question in the evaluation asked whether the exam corresponded to the teaching activities in the course with regards to form, content, and level of complexity.
7 student answered that they completely agreed, 5 that they agreed, 3 that they felt neutral, and 1 that they disagreed.
The average student thus agreed that the exam corresponded to the rest of the activities in the course.
Several students wrote additional comments concerning their confusion with regards to proofs requiring classical proof rules.
One student asked for more problems of ``medium difficulty''.

The next question in the evaluation asked whether the examination form and content corresponded to the learning objectives of the course.
6 students answered that they completely agreed, 5 that they agreed, 4 that they felt neutral, and 1 that they disagreed.
The average student thus agreed that the exam corresponded to the learning objectives of the course.

The final section of the evaluation was for further comments and suggestions.
Several students wrote that they found that the exam matched what they expected, and that the test exam was helpful in this regard.
Some students suggested to remove the exam and instead move to a project-based evaluation of their learning.
One student had issues with figuring out how to correctly abort proof attempts to move on to the next question without getting errors in Isabelle.
We will return to these suggestions in the section on potential future work.

\subsection{Isabelle proofs without automation}
The first problem of the exam seemed to be one of the hardest.
The problem was, as mentioned, split in two questions, and one of them was significantly harder than the other.
The easier question was intended to give students an ``early win'' before getting into the rest of the problems.

While most students completed the easier question, very few students completed both questions.
About half of the students had a reasonable attempt at a proof for the harder question, but were missing significant parts.
Several students did not hand in anything at all for the harder question.

This was essentially as expected, since this type of problem requires creativity and ``getting the right idea'' to solve.
The harder question required the use of classical proof rules, which are in our experience generally confusing to students.

\subsection{Verification of functional programs in Isabelle/HOL}
The second problem of the exam was solved completely by most of the students.
Several students only solved one of the questions fully, and only a few students solved none of the two questions.
This may indicate that we were a bit too cautious with the difficulty of the questions in this problem.
Our experience from similar problems given in the assignments throughout the course is that it is difficult to predict whether students will find program verification problems hard or easy.
We thus decided to err on the side of making the problem too easy.

\subsection{Natural deduction proofs}
The third problem of the exam was solved completely by around half of the students.
This problem was also split in an easier and a harder question, and all of the students solved the easier question.
Several students had a reasonable attempt at a proof for the harder question, but were still missing significant parts.

The situation here is similar to the one in the first problem, and again the harder question required the use of classical proof rules.
This again suggests that students find classical proof rules harder to use.

\subsection{Sequent calculus proofs}
The fourth problem of the exam was solved completely by all of the students.
This may be slightly surprising, since one of the questions in this problem also required the use of classical proof rules.
In the sequent calculus system of SeCaV, however, the students do not have to explicitly choose where in the proof to apply a classical proof rule, in contrast to the natural deduction systems of NaDeA and Isabelle/HOL.
As such, proofs involving classical proof rules are not necessarily harder than proofs which do not when working in SeCaV.
For this reason, we might consider increasing the difficulty of this type of problem in future exams.

\subsection{General proofs in Isabelle/HOL with Isar}
The fifth and final problem of the exam was solved completely by around half of the students.
Most students solved one of the questions completely, and one of the questions partially.
Only a few students did not solve any of the questions, and almost all of these students handed in partial solutions.

One of the questions was harder than the other, so this was as expected.
This problem was also the one where partial solutions were easiest to obtain, since the questions asked students to fix a number of more or less independent errors.

\section{Future work}
There are many potential avenues for further work.
As discussed above, we are continuously adjusting the difficulty of our exams based on our experiences each year.
This necessarily also includes creating new problems for each exam.
This presents an issue, since finding problems of the right difficulty is, in our experience, not very easy.
An interesting potential line of research would accordingly be to develop guidelines and evaluation criteria for creating problems.
We expect that this will not be very easy, since different students may find different problems difficult, and since it is generally difficult to predict how difficult a verification problem is.
Rigorous studies of larger student populations and problem classes would be necessary to design good guidelines.
One possibility would be tracking the behaviour of students while solving problems of certain types to determine the issues they run into, then designing criteria to take the most common errors and misunderstandings into account when estimating the difficulty of a problem.
We expect that new tools would need to be developed to accommodate this kind of behavioural study.

A related potential line of research is the creation of more auxiliary tools to control the potential difficulty of problems.
As detailed above, we have already created several tools which interface with Isabelle, but limit the options the students have, making it easier to estimate the difficulty of the problems we create.
Our method of asking students to finish existing partial proofs is another way to limit the ways in which students can misunderstand the problem.
We expect that there are many other ways which may already be in use at various institutions, and we encourage others to share their ideas and experiences.
Additionally, we expect that experiences from other fields with similar problems, e.g.\ geometry or graph theory, could also be valuable, and conversely, that our experiences could be used to develop auxiliary tools for use in these fields.
It seems that students have a remarkably hard time solving problems which require the use of formal classical reasoning, so constraining the difficulty of these is particularly interesting.

A completely different option is to dispense with a written exam altogether.
We see two main options for evaluating the learning outcomes without a written exam: an oral exam based on assignments handed in during the course, or a larger project with a report to be handed in at the end of the course.
An oral exam works well for relatively small courses, but becomes infeasible as the course grows.
At the scale of our course, we do not believe that an oral exam could be implemented without requiring excessive amounts of time for examinations.
On the other hand, we do believe that a larger project with the objective of proving a number of theorems could be implemented without requiring much more time for grading than a written exam.
Such a project would however need to be structured in a way that allows students to progress in reasonable steps without overwhelming them, and we expect that designing the project assignment would be a large amount of work.
It might also be even harder to estimate the difficulty of such a project in comparison to exam problems.

Another line of potential future work concerns the development of problems that can be graded fully or nearly fully automatically.
As detailed above, Isabelle already makes grading quite easy with our existing types of exam problems.
For very large scale courses, however, fully automated grading may be necessary.
While we could of course require that the files students hand in are free of syntax errors (such that they can be processed by the batch checking tools of Isabelle), we believe that such a restriction may be too severe.
Our experience is that many students have a difficult time distinguishing between errors and correctly aborted proof attempts, and thus hand in files containing plenty of errors.
One possibility might be to develop a tool for fixing Isabelle theories with errors, either automatically or by suggesting fixes for any errors found in the file.
We do not expect that a practically usable version of such a tool would be very easy to develop.

In any case, if we were certain that all student submissions could be processed by the Isabelle batch checking tools, fully automated grading would still require careful design of the problems to ensure that they contain sensible milestones that can be graded individually.
Inspiration for this might be taken from the automated grading scripts for the Software Foundations books, which contain exercises in the Coq proof assistant \cite{Pierce:SF1, Pierce09}.
We expect, however, that designing questions with reasonable milestones for our problems concerning natural deduction proofs and sequent calculus proofs would require a rework of our tools to support stopping partway through a proof.
We currently also deduct a few points for very bad style, in the sense that proofs which are significantly longer or much more complicated than they need to be do not give full marks.
We expect that automatically performing such a judgment would be difficult, but it might be possible to develop a tool which fully automatically grades reasonable proofs and can mark ``problematic'' submissions for manual review.

\section{Conclusion}
When conducting a course on automated reasoning using Isabelle, we need to test learning outcomes, and have chosen to do so in an exam setting.
We have described our approach for doing this, and explained how Isabelle can help us quickly grade exam submissions.
We have also explained the various kinds of questions in our exam, and how their design tests various aspects of student understanding of logic in general, and of formal proofs in Isabelle/HOL in particular, through a number of examples.

Students generally seem to understand how to solve the problems and are positive about the exam format.
The main challenge in designing exams using this approach is to balance the difficulty appropriately, avoiding trivial problems while still allowing all students to make at least some progress in each problem.
We have explained how simple programming problems can be combined with problems about completing larger proofs to test student understanding at several levels.
Other courses may have different design constraints, especially those with many students, where fully automated grading may be desirable or necessary.

We believe that our experiences and the techniques described in this paper may also be useful for exams in other fields with similar constraints and a desire or need to fully or partially automate grading.
Such fields might include geometry, graph theory, and any other field in which reasoning plays a significant role and where students may easily misunderstand questions or proofs.
We expect that bespoke tools similar to NaDeA and SeCaV would be necessary to make formal reasoning in these fields accessible to students without previous experience with proof assistants.

\section*{Acknowledgements}
We would like to thank Asta Halkjær From, Deniz Sarikaya, Frederik Lyhne Andersen, Simon Tobias Lund, and the anonymous reviewers for their helpful comments on drafts.

\

\bibliographystyle{eptcs}
\bibliography{references}

\end{document}